# Robust Optical Pulling of Plasmonic, Chiral, and Dielectric Nanorods in Single and Cluster Configurations Using Hyperbolic Metamaterials

Md. Al Amin Chy[1], Md. Ajijul Haque Sakib Opu[1], Md Fujael Ahmed[1], Rashedul Albab[1], Md. Janibul Alam Soeb[*]

[1]Department of Electrical and Electronics Engineering, Sylhet Engineering College, Sylhet-3100, Bangladesh

[*]Department of Farm Power and Machinery, Sylhet Agricultural University, Sylhet-3100, Bangladesh

**Abstract**

Optical manipulation enables the control of microscopic particles using light and has become an active area of research in recent years. One particularly interesting aspect is the possibility of generating optical pulling forces with hyperbolic metamaterials (HMMs), which leads to unusual light–matter interactions. In this work, the pulling response of plasmonic, chiral, and dielectric nanorods is studied on an HMM substrate. The system is illuminated by a time-harmonic, plane-polarized laser beam propagating along the +Z to –Z direction. The analysis is carried out under two different configurations. In the first case, plasmonic, chiral, and dielectric nanorods are arranged in a heterogeneous cluster over the HMMs and analyzed for the optical force response under three environmental conditions: fully immersed in water, partially immersed (half air–half water), and fully in air. Full-wave simulations reveal that in all three environments, all the objects (plasmonic, chiral, dielectric) nanorods consistently experience an optical pulling force at the $10^{-20}$ N scales for clustered configuration. In the second case, each nanorod type was independently placed on the same HMMs substrate within a full-air medium and investigated their force behaviors, while the plasmonic experience 100 times more than others for the single setup, which is $10^{-19}$ N ranges. Like the case of the cluster, three nanorods alone demonstrate a stable optical pulling force, also getting the pulling force with the variation of length, radius, and distance between them, which makes this optical setup more practical. Plasmonic behavior changes a lot with the arrangement of the nanostructures. On the other hand, dielectric nanorods are affected more by their length than their position. In our results, optical pulling appears for all the nanorods even with a simple HMM structure. This suggests the approach works in a stable way. It may also be useful for scaling the optical manipulation of anisotropic nanostructures, especially for sensing and nanophotonic applications.

## 1. Introduction

Study of optical forces is a crucial part of research on the interaction between light and matter. It helps us understand the role of momentum carried by light and transfer of forces to physical objects. Many scientific experiments have been performed with the help of manipulation methods in optics [1–5] . In this article, an important contribution to the field of optics made by Ashkin's fundamental research has significantly affected many scientific studies and applications [6,7] . By exploring optical forces, researchers can understand the interaction of light and matter, which opens up new ways of innovative utilization in various areas. Currently, the use of optical manipulation is mandatory in many fields such as mechanics, biology, quantum physics, etc. [8] based on the principles of radiation pressure of electromagnetic waves. Optical manipulation has different methods, and one such method known as "optical pulling" has attracted considerable attention because of its unique characteristics [9–11]. This research [12,13] aims at studying the physics involved in the process of optical pulling forces. The latter are associated with the electromagnetic force. In turn, the optical scattering force involves pushing of the object along the direction of propagation of light. The optical pushing force [14,15] can be referred to as the pulling force operating in the reverse direction, and the result is movement of an object towards the light source. Thus, it was possible to extend the area of application of the force applied to objects for their manipulation. For achieving optical pulling forces, scientists use different techniques such as increasing linear momentum of light, shaping and modifying properties of the object and the medium in question [16,17]. However, due to localized light, manipulation of nanoparticles poses certain challenges. Metamaterials offer greater opportunities for manipulating nanoparticle objects [18–21]. This is what makes hyperbolic metamaterial [22] more effective, since it allows having highly confined modes and high density of states. Such unique features of a hyperbolic material are attributed to hyperbolic dispersion, causing negative permittivity in the principal crystallographic directions [23]. The hyperbolic metamaterials, which constitute the basic building blocks of hyper-crystals, are known as those materials whose dielectric permittivity differs along two orthogonal directions. The strong anisotropy property in such materials enables very high wavevector ($\kappa$) bulk waves that are theoretically unbounded and define the hyperbolic dispersion relation [24]. Studying optical forces in the context of hyperbolic metamaterials is a fascinating study. Several methods have already been used for constructing an optical tractor beam (OTB) [25–28]; however, these methods are rather complicated. Optical manipulation has made great stride over the last few years. To the best of our knowledge, no study has yet proposed a way to create a single optical beam configuration in which plasmonic, chiral, and dielectric types of nanorods experience optical pulling force in both single and clustered form. Exclusion or extraction of microplastics (dielectric) from a clustered configuration is a costly approach. Plasmonic (Gold) Nanorods are costly as well [29]. As such, an easy optical configuration plays an important role in attracting various kinds of materials rather than using costly chemicals and difficult methods. Hence, this paper presents a hyperbolic metamaterial substrate and time harmonic plane-polarized laser beam (incident light) to create an optical pulling force on plasmonic, dielectric, and chiral nanorods. Hyperbolic metamaterials are a type of combination between dielectric-metal or metal-dielectric multilayers that have high anisotropic properties with unique phenomena including negative refraction [30–32] , anomalously cavity resonance [33], super-regulation imaging [34–36] and topological transition [37,38]. In addition, hyperbolic metamaterials display an isofrequency [39] effect to create an optical pulling force from an enhancement of forward optical momentum [40]. New progress in fabricating nanostructures results in vertical hyperbolic metamaterials [41]. In this paper, a bilayer structure of 4 periodic plasmonic and dielectric hyperbolic metamaterials is studied, where it is observed that the pulling force is created in heterogeneous clusters through the negative gradient force and electric dipole-quadrupole interaction on dielectric nanorods ($SiO_2$). For the case of plasmonic materials, optical pulling force is generated by the current density behavior reversal while the optical pulling force is generated in the case of chiral through the interference of electric magnetic dipoles. All nanorods experience optical pulling force when they are placed in a single cluster configuration.

The use of any other type of material to generate OTB till now has not yet been proposed [40]. Moreover, no idea has ever been provided in the previous literature on an optical setup which can produce an optical tractor beam on either the cluster level or the single nanorod level. In this paper, we make our first attempt to develop the optical setup (both cluster and single) of chiral, plasmonic, and dielectric nanorods.

## 2. Methodology

This work focuses on analyzing the direction of optical forces acting on both an individual nanorod and a small group of nanorods (NRs), including dielectric, plasmonic, and chiral types, when they are positioned on a hyperbolic metamaterial under different surrounding environments such as air, water, and their combined medium. The overall aim of our research work is to construct a general optical system through which the phenomenon of OTB may be observed for all three objects (i.e., dielectric, plasmonic, and chiral) placed above our designed system. We have placed the dielectric ($SiO_2$), plasmonic (Au), and chiral nanorods above the hyperbolic metamaterial structure, as depicted in Fig. 1 and Fig. 2(a(i)). To start our investigation, the plasmonic nanorod was placed centrally with respect to the substrate structure, while the other two nanorods were placed on either side of the plasmonic nanorod,

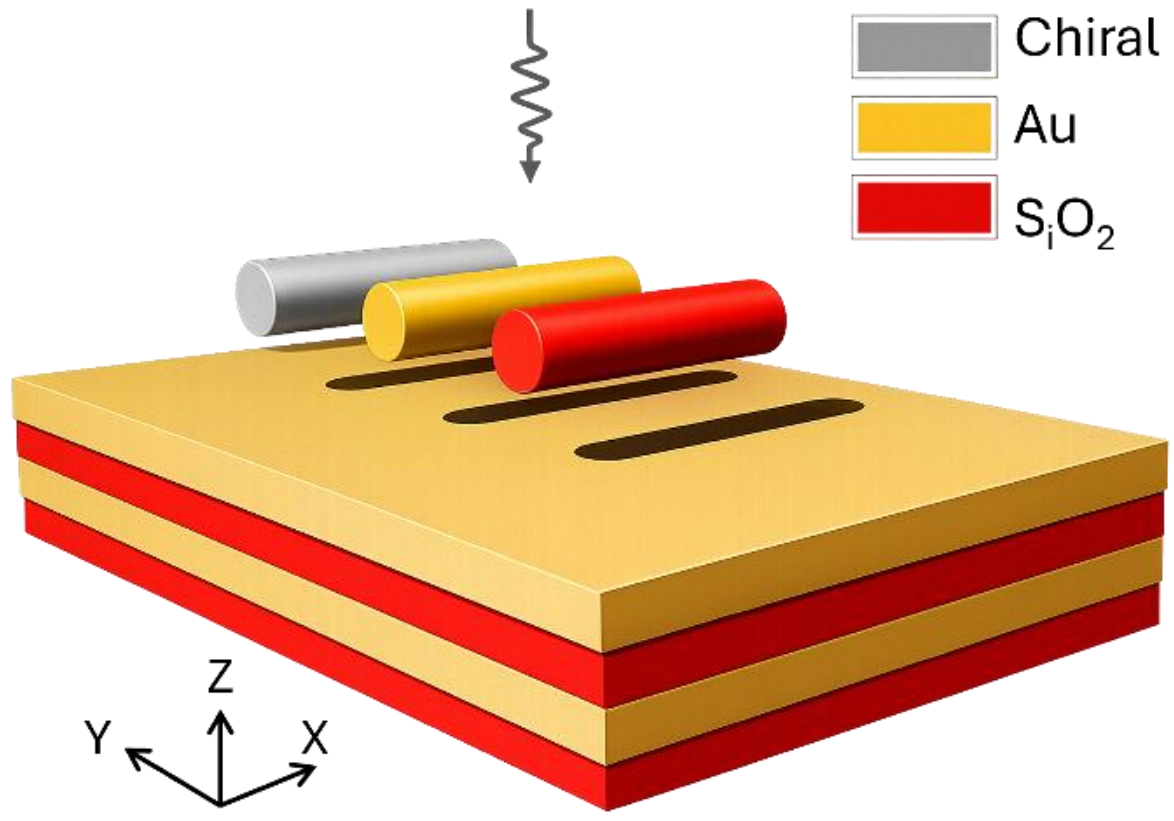

**Fig. 1** illustrates our proposed optical setup being kept on the right-hand and left-hand sides, respectively. But it is not realistic to keep all nanorods static within the structure. Thus, different positions were considered for NRs in simulations. For the proposed optical configuration, a plane-polarized time-harmonic laser beam is considered as the excitation source. The incident wave is non-structured in nature and exhibits magnetic field polarization along the Y-axis. The beam propagates from the +Z direction toward the −Z direction, while maintaining normal incidence with an orientation angle of θ = 0° relative to the normal of the surface. Also, the expression to calculate the wave function of light being used can be calculated through the formula:

$$H = H_o e^{jkz} \hat{y} \tag{1}$$

In addition, the present investigation is further extended to examine several HMM-based optical configurations under different surrounding environments. These configurations include a partially immersed arrangement consisting of half-air and half-water regions, as illustrated in Fig. 2(b(i)), as well as a fully water-submerged configuration shown in Fig. 2(c(i)). For the water medium considered in this analysis, the refractive index is assumed to be n = 1.33. Note also that the simulation condition other than background medium remained the same as those in full air medium. We have used the plasmonic-dielectric bilayer type hyperbolic metamaterial in our study, where both layers have dimensions of 600 nm in length and width respectively while thickness has been assigned the value of t = 20nm and the period of HMM denoted as 'p' is 40nm. The distance between inter-nanorod gap between the NRs in our simple cluster has been chosen as d = 25nm, while all the nanorods utilized in our simulations have radius of R = 30nm and length of l = 150nm. The cluster is located at height h = 30nm above the HMM. The selected HMM forms the concave type of isofrequency contour by choosing appropriate photonic crystal parameters, which help in enhancing the forward linear momentum of the field dispersing from the elliptical material placed within the photonic crystal [42]. In this way, because of fundamental plane waves rather than structured waves of beam profile, the objects can be displaced by means of generated optical pulling forces. The strong optical pulling force acting on the HMM structure opens an interesting avenue for manipulating optical objects in nanoscale dimensions. For the HMM structure consisting of plasmonic dielectric multilayers, the components of effective permittivity along $\boldsymbol{\varepsilon}_{||} = \boldsymbol{\varepsilon}_z$ and across ($\boldsymbol{\varepsilon}_{\perp} = \boldsymbol{\varepsilon}_x = \boldsymbol{\varepsilon}_y$) 'z' axis are given by [39]:

$$\begin{cases} \varepsilon_{||} = \frac{\varepsilon_d \varepsilon_m}{f\varepsilon_d + (1-f)\varepsilon_m} \\ \varepsilon_{\perp} = \varepsilon_m f + (1-f)\varepsilon_d \end{cases} \tag{2}$$

Here, $\boldsymbol{\varepsilon}_d$ and $\boldsymbol{\varepsilon}_m$ represent the permittivities of the dielectric and plasmonic material, respectively. The parameter $f$ denotes the filling ratio and is defined as $f = t/p$, where it expresses the thickness proportion of the plasmonic–dielectric layers within a single period. Additional details regarding the proposed HMM structure are provided in Supplementary Note 1. The constitutive relations associated with the chiral nanorod material are expressed as follows [43,44]:

$$\mathbf{D} = \varepsilon_0 \varepsilon_c \mathbf{E} + i\varsigma\sqrt{\boldsymbol{\varepsilon}_0 \mu_0}\mathbf{H} \tag{3}$$

$$\mathbf{B} = \mu_0 \mu_c \mathbf{H} - i\varsigma\sqrt{\boldsymbol{\varepsilon}_0 \mu_0}\mathbf{E} \tag{4}$$

In these relations, $\boldsymbol{\varepsilon}_c$ and $\mu_c$ denote the permittivity and permeability of the right-handed chiral medium [$\boldsymbol{\varepsilon}_c$ = $(1.45)^2$ and $\mu_c = 1$ ], respectively. The chirality parameter, $\varsigma$, equals +1 in our case, and fulfills the following relation: $\varsigma^2 < \varepsilon_c \mu_c$ [43]. $\varepsilon_0$ and $\mu_0$ stand for the permittivity and permeability of free space, respectively. In the numerical simulations considered, the complex refractive index of plasmonic (Au) material, utilized for both plasmonic nanorods and plasmonic layers in HMM, is determined based on experimental results presented by Johnson [45]. The refractive indices for dielectric ($SiO_2$) nanorods, used for the dielectric layers of HMM, are equal to n = 1.45. In the full-wave simulations conducted using COMSOL Multiphysics 6.0 [46], the incident wavelength λ is swept from 700 nm to 1600 nm with an interval of 5 nm. This wavelength-dependent analysis is carried out to determine the optical forces exerted on the three nanorods.

For the three surrounding background configurations considered in this study, Fig. 2(a) corresponds to Case 1, representing the air medium. Figure 2(b) is defined as Case 2, where the structure is partially surrounded by air and water, while Fig. 2(c) represents Case 3, corresponding to the fully water-immersed environment. In the full-wave electromagnetic analysis, the optical forces evaluated outside the nanorod region are termed exterior optical forces, whereas the forces calculated within the nanorod domain are identified as interior optical forces. The exterior force components are obtained from the surrounding background fields by integrating the time-averaged Minkowski stress tensor over a circular boundary of radius 'a', positioned 3 nm away from the scatterer [47,48].

$$\langle \mathrm{F}_{\mathrm{Total}}^{\mathrm{out}} \rangle = \oint [\langle \overline{\overline{\mathbf{T}}}^{\mathbf{out}} \rangle] \bullet ds \tag{5}$$

Where, $\langle \overline{\overline{\mathbf{T}}}^{\mathbf{out}} \rangle$ may be written as,

$$\langle \overline{\overline{\mathbf{T}}}^{\mathbf{out}} \rangle = \frac{1}{2} \Re [(\mathbf{D_{out}} \mathbf{E_{out}^{*}} + \mathbf{B_{out}} \mathbf{H_{out}^{*}}) - \frac{\mathbf{1}}{\mathbf{2}} \overline{\overline{\mathbf{I}}} (\mathbf{E_{out}^{*}} \bullet \mathbf{D_{out}} + \mathbf{H_{out}^{*}} \bullet \mathbf{B_{out}})]$$

The subscript "out" denotes the total external field associated with the scatterers, which includes both the incident and the scattered components. The electric field, electric displacement field, magnetic field, and magnetic flux density are represented by E, D, H, and B, respectively. In addition, the notation $\langle \ \rangle$ is used to indicate time averaging, while $\overline{\overline{\mathbf{I}}}$ represents the unit tensor. Comparable computational approaches have been employed in earlier studies [49,50] to analyze the optical forces acting on various objects.

## 3. Results and Discussion

Linearly polarized time-harmonic plane waves with different optical set-ups were observed in this study in three distinct nanorods. In a recent study, distinct Rayleigh particles were utilized to create an optical pulling set-up [51]. On the contrary, ref [52] reports that optical pulling force was not observed on these particles, which implies that dielectric particles exhibit optical pulling force, while plasmonic and chiral particles exhibit optical pushing and lateral force, respectively. Observation of these optical forces starts with numerical computation of clustered configurations followed by single configurations. Optical pulling forces observed on distinct nanorods having chiral, plasmonic, and dielectric materials with three different environments, such as

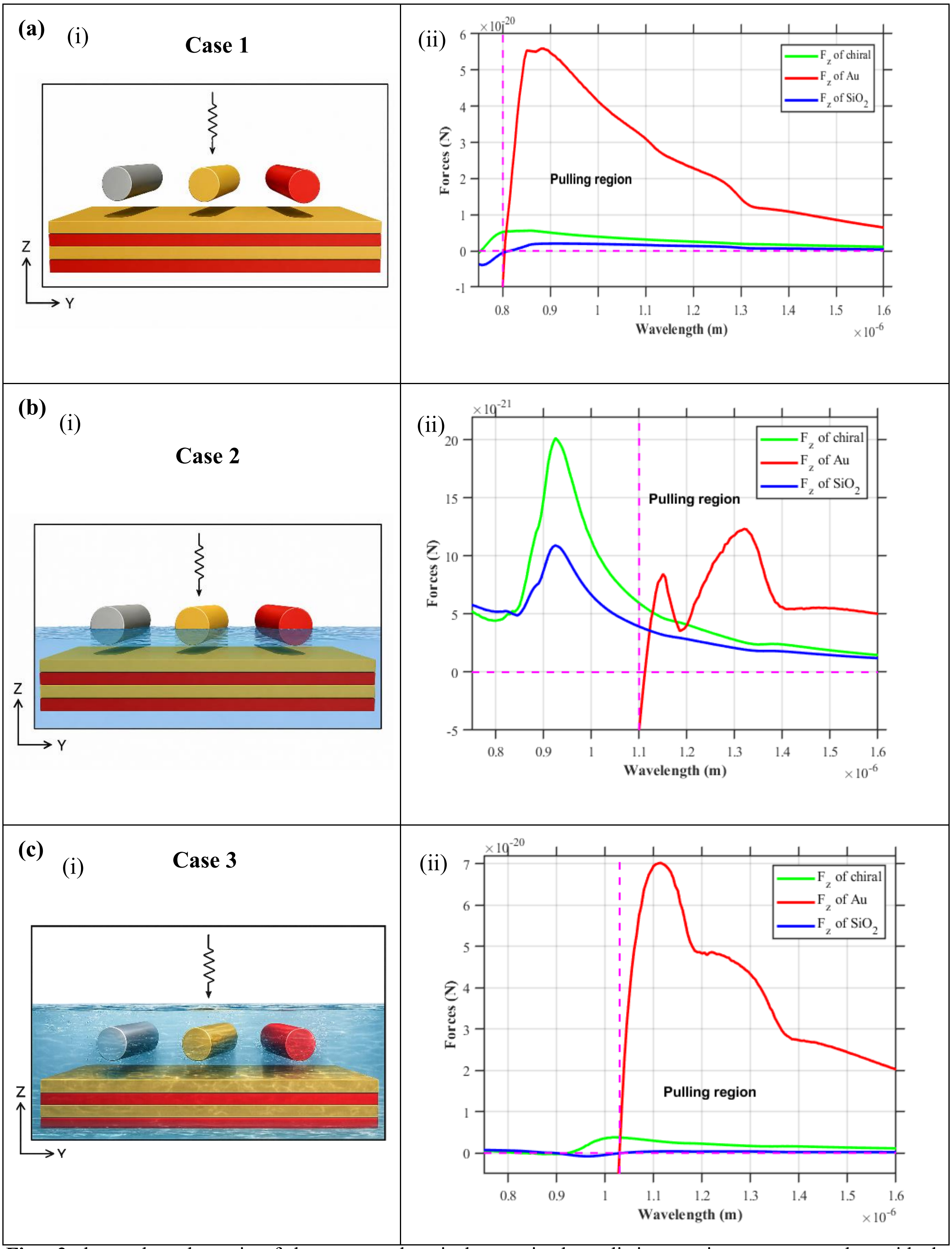

**Fig. 2** shows the schematic of the suggested optical setup in three distinct environments, together with the accompanying full wave simulation results. Case 1 (full air medium) is illustrated in (a), the setup is schematically shown in (i), and the full-wave simulation results for our suggested Case 1 configuration are shown in (ii). In fig.(b), A half-submerged water medium illustration of Case 2, (i) a schematic of the setup, and (ii) the full-wave simulation results for the configuration suggested for Case 2. In fig.(c), A schematic of the setup, (i) an illustration of Case 3 (completely submerged in water), and (ii) the outcomes of the full-wave simulation for our suggested Case 3 configurations.

air, half-air & half-water, and water medium, considered in cases 1,2, and 3, respectively. For achieving an OTB, the hyperbolic metamaterial is used as a substrate in this proposed approach. In Fig. 2(a), where in the optical cluster was encircled by an air medium, we did this with the help of hyperbolic metamaterial in our setup. Multiple reflections from the reflected beam due to NRs were prevented through a homogeneous background and substrate removal. However, as shown in Fig. 2(a(ii)), the full-wave simulation results show clearly that all the three nanorods undergo optical pulling force starting at 800 nm, and this goes on for wide wavelength ranges (1600 nm). It is observed that optical pulling force appears for the half-air-half-water medium from 1120 nm onwards (Fig. 2(b)) while the OTB effect is observable from 1030 nm onwards in the submerged water medium (Fig. 2(c)). It can be seen that the pulling effect has continued for up to 1600 nm in all these three scenarios.

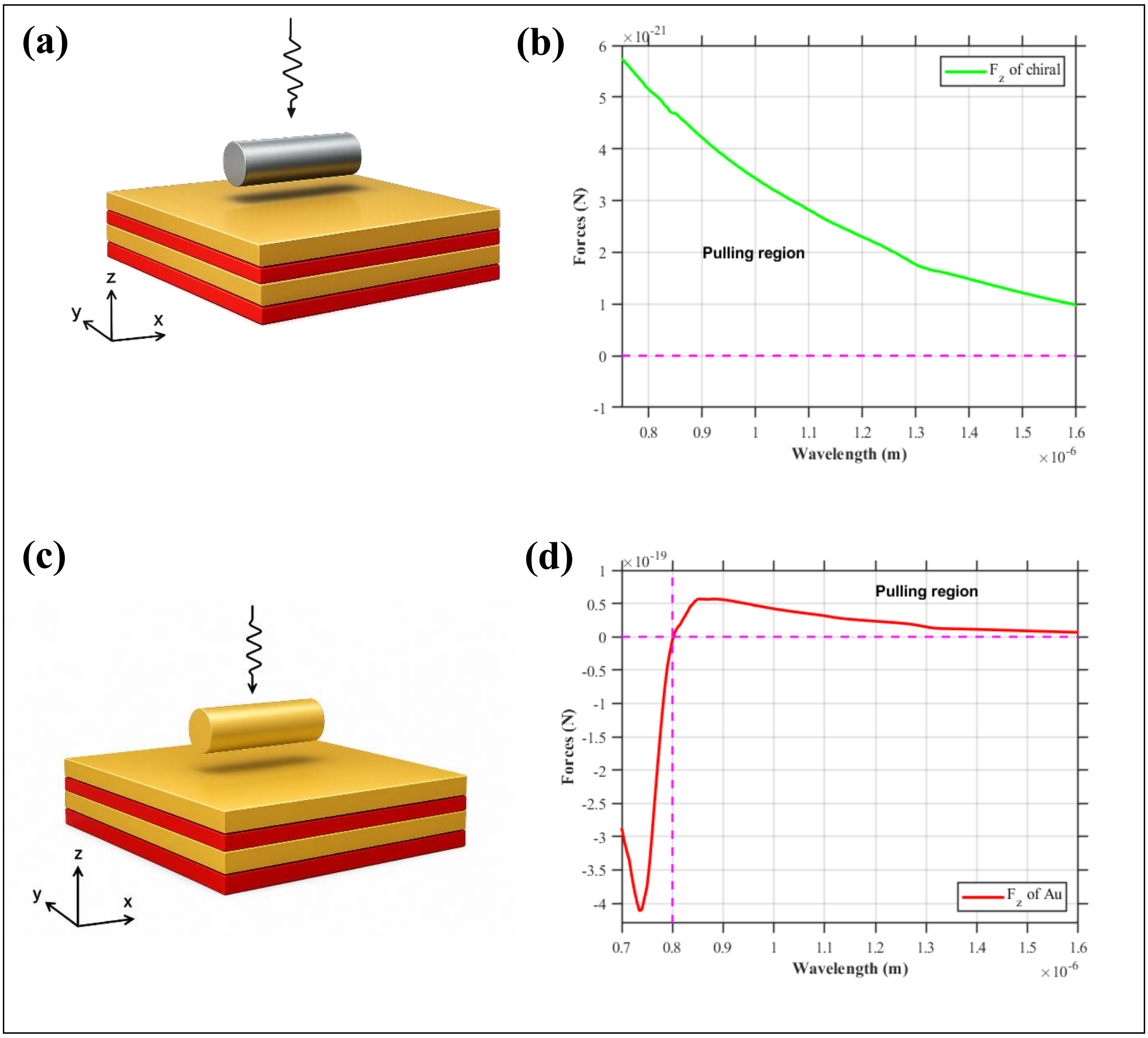

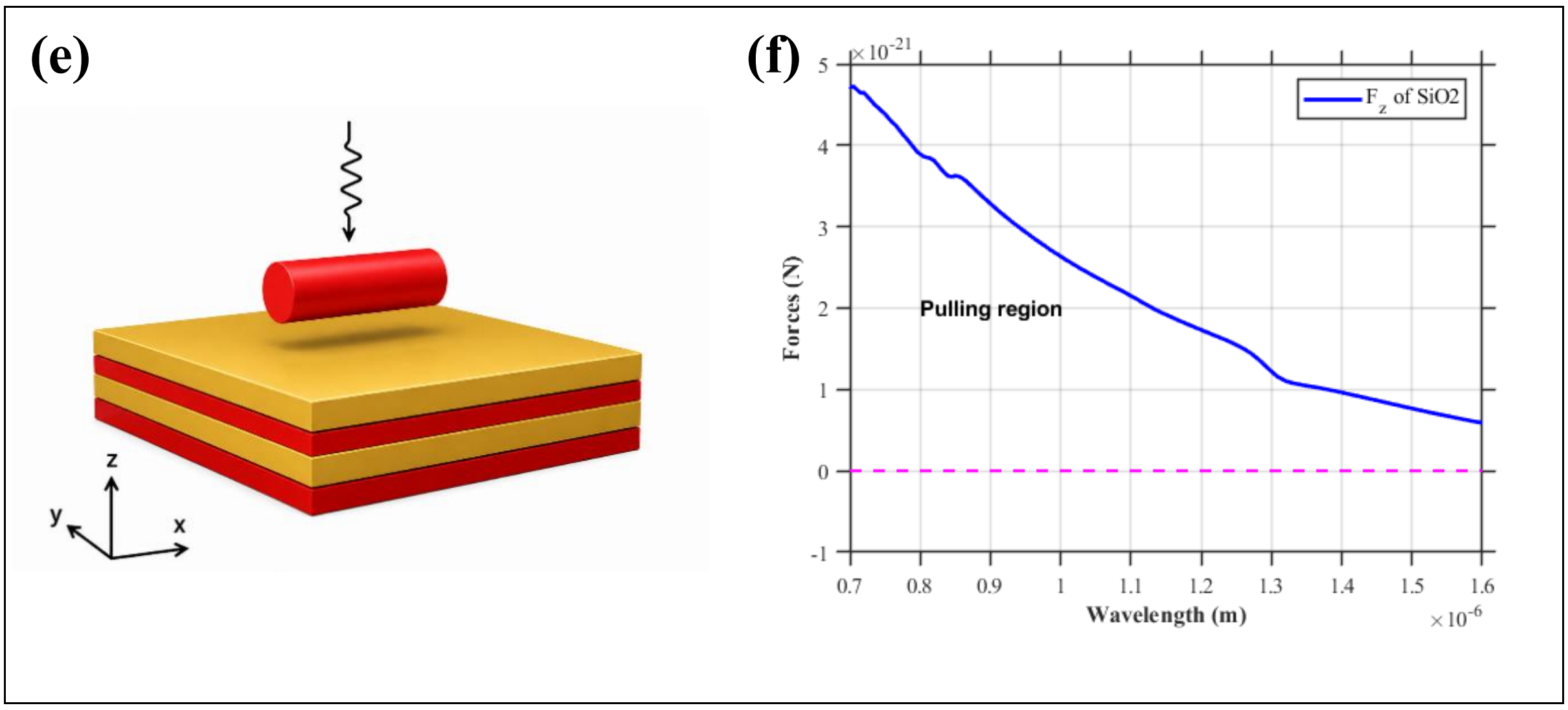

**Fig. 3** illustrates the schematic representation of the proposed single optical setup: (a) depicts the chiral nanorod configuration, (b) presents its corresponding full-wave simulation results, (c) shows the plasmonic nanorod, (d) provides the full-wave simulation outcomes for the plasmonic case, (e) represents the dielectric nanorod, and (f) displays the associated full-wave simulation results.

Having observed the cluster configurations in diverse mediums, the research work moved further into examining single configurations. In addition to this, we have also shown that in case of individual nanorods of either chiral, plasmonic, or dielectric nature as shown in Fig. 3, There is an occurrence of optical pulling force when placed individually on our proposed HMM-based substrate. According to Fig. 3(b & f), optical pulling can occur for both chiral and dielectric nanorods ranging from 700 to 1600 nm in total duration of observations. Also, it is observed that there occurs optical pulling of our system in the presence of plasmonic nanorod from 800 to 1600 nm, as shown in Fig. 3(d). It needs to be noted that optical pulling was also observed for our system ranging between 800 and 1600 nm in presence of all the three nanorods arranged in clusters in our optical setup. It shows that our data shows similar pulling domains for both the scenarios. It also needs to be stated that in all these optical set-ups, it was found that optical pulling force was achievable when the setup consisting of a single nanorod was kept in different mediums, such as completely submerged in water or half-submerged in water. To create a practical optical setup for obtaining optical pulling forces of nanorods, observations were made of all possible observations in this research work. A complete mechanism applicable to all NRs in every possible scenario is elaborated in our Supplementary Note 2. The subsequent subsection gives a detailed explanation of the mechanisms applicable to all NRs.

### 3.1 Mechanism of optical pulling in a chiral nanorod

It is well known that the chiral materials have a tendency towards light due to their chirality [53,54]. Here, the constitutive relation [55] gives the dependence between electric and magnetic momentum "p" and "m".

$$P = \alpha_e \mathbf{E} + i_\chi \mathbf{H} \tag{6}$$

$$m = -i_\chi \mathbf{E} + \alpha_m \mathbf{H} \tag{7}$$

In this context, $\alpha_e$ denotes the complex electric polarizability, $\alpha_m$ refers to the magnetic polarizability, and $\chi$ indicates the complex chiral polarizability. To analyze the optical pulling force exerted on a chiral nanorod (NR) placed above the proposed optical configuration, it is necessary to consider the influence of multipolar excitations generated inside the scatterer. Earlier research has shown that, since these structures are much smaller than the wavelength of the incoming light, the dominant contributions arise primarily from electric and magnetic dipole moments. The subsequent discussion aims to clarify the behavior of the optical pulling force acting on the chiral NR. Based on Refs. [54,56], the total time-averaged optical force can be formulated in terms of the polarizability parameters as follows:

$$\langle \mathbf{F} \rangle = \langle \mathbf{Fe} \rangle + \langle \mathbf{Fm} \rangle + \langle \mathbf{Fe} - \mathrm{m} \rangle \tag{8}$$

$$\langle \mathbf{Fe} \rangle = \frac{1}{4}\Re(\alpha_e)\nabla|\mathbf{E}|^2 + \frac{k}{2n}\Im(\alpha_e)\Re(\mathbf{E}\times\mathbf{B}^*) + \frac{1}{2}\Im(\alpha_e)\Im[(\mathbf{E}^*\bullet\nabla)\mathbf{E}] \tag{9}$$

$$\langle \mathbf{Fm} \rangle = \frac{1}{4}\Re(\alpha_m)\nabla|\mathbf{B}|^2 + \frac{k}{2n}\Im(\alpha_m)\Re(\mathbf{E}\times\mathbf{B}^*) + \frac{1}{2}\Im(\alpha_m)\Im[(\mathbf{B}^*\bullet\nabla)\mathbf{B}] \tag{10}$$

$$\langle \mathbf{Fe} - \mathrm{m} \rangle = -\frac{k^4}{3}\sqrt{\frac{\mu}{\epsilon}}\Re(\alpha_e\alpha_m{}^*)\Re(\mathbf{E}\times\mathbf{B}^*) - \frac{k^3}{3}\mu\Im(\alpha_e\alpha_m{}^*) \times \frac{1}{2}\{\nabla|\mathbf{E}|^2 - \Re[(\mathbf{E}^*\bullet\nabla)\mathbf{E}]\} \tag{11}$$

**F**m, **F**e, and **F**e − m are the magnetic dipole, electric dipole, and recoil force, respectively, in the above (9)-(11) equations, which one can be written as like,

$$\mathrm{F_e} = \frac{1}{2}\Re\{(\nabla\mathbf{E}^*)\bullet\,\mathbf{p})\}$$

$$\mathrm{F_m} = \frac{1}{2}\Re\{(\nabla\mathbf{B}^*)\bullet\,\mathbf{m})\}$$

$$\mathrm{F_{e-m}} = -\frac{k^4}{12\pi\epsilon_0 c}\Re\{\mathbf{p}\times\mathbf{m}^*\}$$

The coupling between the electric and magnetic dipoles, represented by the third term of Eq. (8), plays the dominant role in producing the optical pulling force observed in the chiral structure. In contrast, the nanorod may experience a conventional pushing force when momentum transfer occurs only through the interaction between isolated electric and magnetic

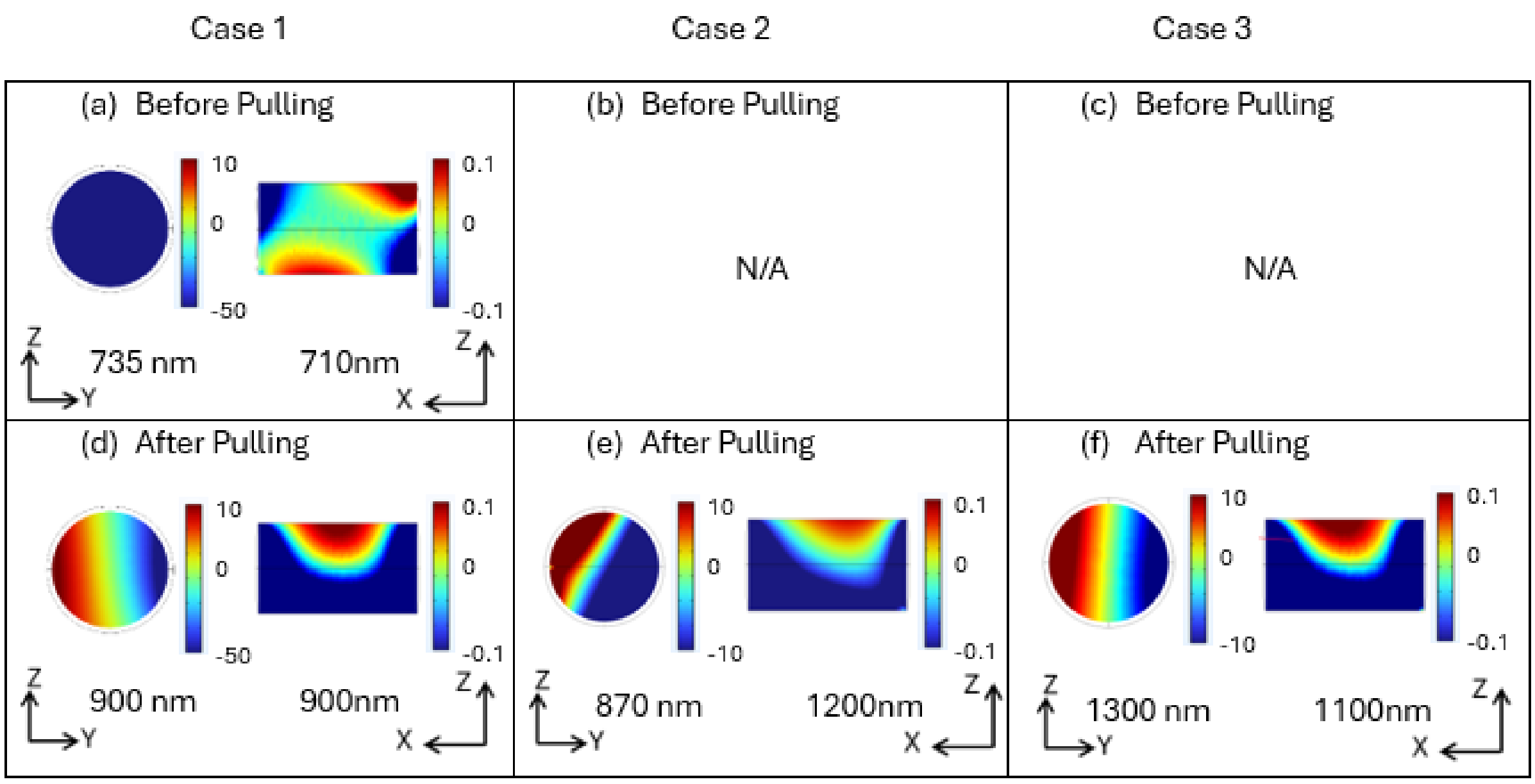

**Fig. 4** illustrates the spatial distributions of the electric and magnetic fields generated by the chiral scatterer for several excitation wavelengths and surrounding media environments. Figures 4(a) and 4(d) correspond to the proposed optical arrangement in air (case 1), where the excitation wavelengths are 735 nm, 900 nm, and 710 nm. For case 2, the same chiral structure is considered in a partially water-immersed condition, and the resulting field responses at 870 nm and 1200 nm are presented in Fig. 4(e). The fully immersed configuration in water (case 3) is shown in Fig. 4(f), where the system is examined at excitation wavelengths of 1300 nm and 1000 nm. Because no force contribution other than the pulling region is observed, the related force values become negligible in Figs. 4(b) and 4(c). In these field maps, the Y–Z and Z–X planes describe the spatial profiles of the electric field intensity (V/m) and magnetic field intensity (A/m), respectively.

dipoles. The electric- and magnetic-field profiles presented in Figs. 4(a) and 4(d) are analyzed for the air-background configuration (Case 1). At the wavelengths of 735 nm, 710 nm, and 900 nm, the chiral nanorod cluster does not exhibit any electric- or magnetic-dipole resonance behavior. Before the appearance of the optical pulling region, the internal field response of the cluster remains weak, indicating the absence of pronounced dipolar excitation modes. Once the chiral nanorod begins to experience optical pulling, clear excitation of both dipole components becomes visible, particularly in Fig. 6(d). For the generation of a stable optical pulling force, the contribution associated with $F_{e-m}$ [54,56,57] must become sufficiently strong, which requires the simultaneous excitation of electric and magnetic dipoles as predicted by Eq. (8). The interaction between these induced dipoles produces a recoil force that further modifies the intensity-gradient and scattering-force components [57]. The field profiles presented in Figs. 4(e), 4(d) and 4(f) confirm the presence of coupled electric and magnetic dipole modes within the pulling region. Therefore, it can be inferred that the unusual pulling behavior of the chiral nanorod in the proposed configuration mainly originates from the concurrent induction of electric and magnetic dipoles inside the structure. In addition, analysis of Figs. S6 and S7 (Supplementary Note 3) suggests that the HMM layer also assists in facilitating the OTB response of the chiral nanorod.

### 3.2 Mechanism of optical pulling in a plasmonic nanorod

It is important for us to understand how momentum contributions to the charges in the plasmonic nanorod work to be able to understand non-conservative optical pulling forces. First, one part of these pulling forces comes from the forces acting on bound charges and currents,

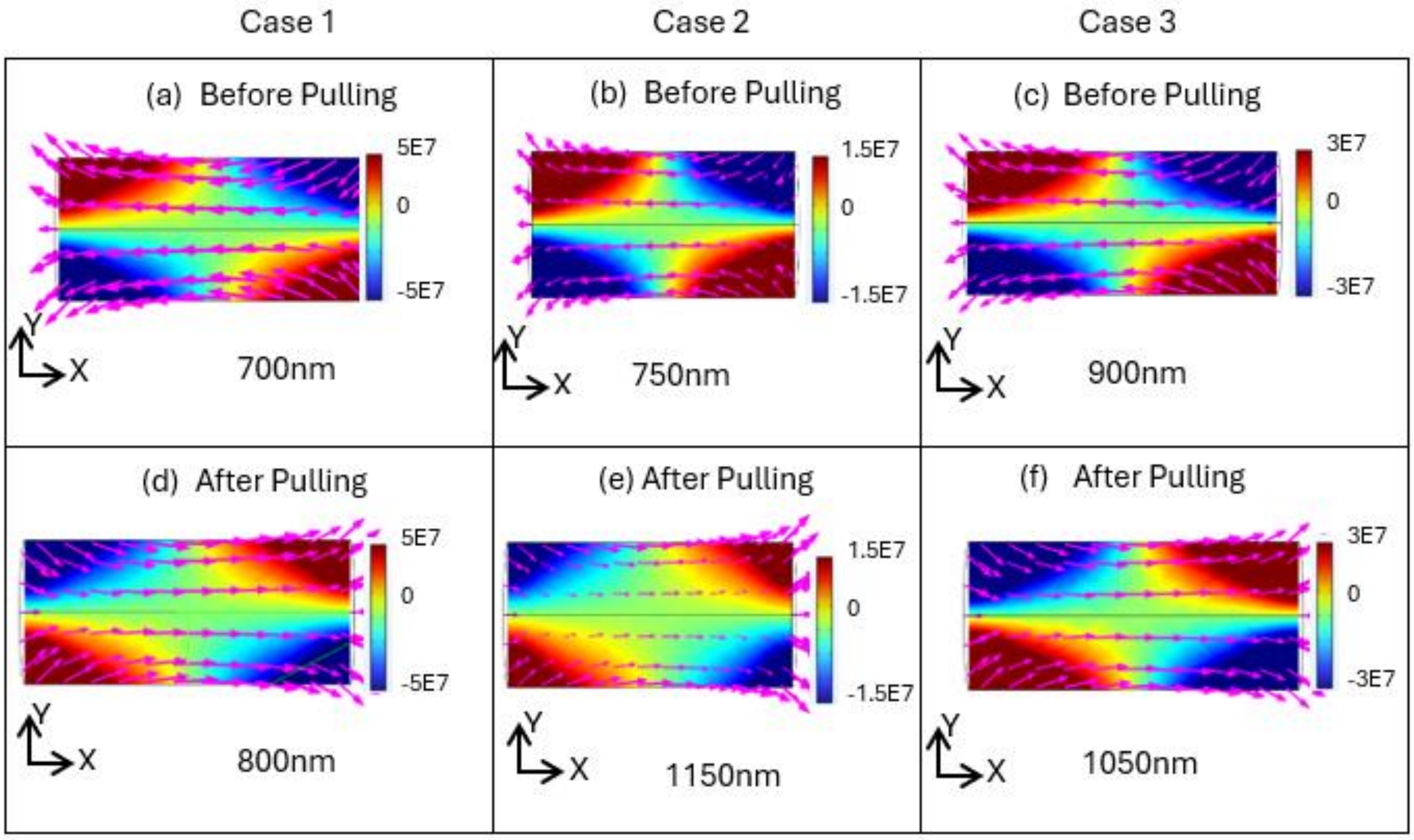

**Fig. 5** illustrates the current-density distributions (A/m$^2$), including both field maps and vector-arrow representations, for the plasmonic nanorod under different excitation wavelengths and environmental conditions. In the air medium (case 1), the responses corresponding to incident wavelengths of 700 nm and 800 nm are presented in Figs. 5(a) and 5(d), respectively, using the optimized optical configuration. For the partially immersed condition in water (case 2), Figs. 5(b) and 5(e) demonstrate the behavior of the same plasmonic scatterer at excitation wavelengths of 750 nm and 1150 nm. Meanwhile, Figs. 5(c) and 5(f) correspond to the fully submerged configuration in water (case 3), analyzed at incident wavelengths of 900 nm and 1050 nm, respectively. The displayed vector directions together with the current-density field distributions clearly indicate the reversal behavior within the plasmonic structure.

whereas the other part is based on the forces acting on free currents [58]. The latter usually have no impact on the analysis of the problem since they generate positive forces [59]. Naturally, the incidences of the SPP waves shown in Figure 2 create pulling forces in the -z direction since this is the direction of propagation of the waves. However, to push the plasmonic scatterers in the required direction, it becomes necessary to impart momentum to the free currents. At the same time, the distribution of induced currents' densities in an active plasmonic device affects the force contribution of the bound charges and currents. Comparing Figures 5a and 5d, one may note that force inversion takes place in the region of current density inversion inside the nanorod made of gold. The same can be concluded by considering the relationship between Figures 5b and 5e, and for Figures 5c and 5f, according to which current density inversion happens only in the case of switching from pushing to pulling forces acting on the scatterers. This happens due to constant current density before force reversal, followed by the inversion near force reversal and maintaining the state in the region after the inversion. It is important to note that the forces exerted on bound charges and currents may take either positive or negative values. A positive force results in a pushing effect, whereas a negative force leads to a pulling action, commonly referred to as an optical tractor beam, which draws plasmonic scatterers toward the source of illumination [60]. The distributions of the Lorentz forces within plasmonic objects may also help in explaining this force inversion [61].

$$\langle F_{total} \rangle = \frac{1}{2} \mathrm{Re}\left[\oint \left(-\frac{1}{2}|E_{Surf}|^2 \nabla\varepsilon\right) ds\right] + \frac{1}{2}\mathrm{Re}[\int(\rho_{free}E_{in} + J_{free}(in) \times \mu_0 H_{in})\ dv] \tag{12}$$

Where, the subscripts "surf" and "in" denote the surface region and the internal region of the scatterer, respectively. By inserting the expression for $F_{total}$ into Eq. (5), Eq. (12) can be reformulated as follows:

$$\oint \langle \mathbf{T^{out}} \rangle \cdot ds = \mathbf{F}_{total} = \langle F_{Mink.}^{Surface} \rangle + \langle F_{Mink.}^{Bulk} \rangle \tag{13}$$

Where,

$$\langle F_{Mink.}^{Bulk} \rangle = \frac{1}{2}\mathrm{Re}[\int(\rho_{free}E_{in} + J_{free}(in) \times \mu_0 H_{in})\ dv]$$

$$\langle F_{Mink.}^{Surface} \rangle = \frac{1}{2} \mathrm{Re}\left[\oint \left(-\frac{1}{2}|E_{Surf}|^2 \nabla\varepsilon\right) ds\right] \tag{14}$$

When no internal free charge density ($\rho_{free}$) or free electric current ($J_{free}$) exists inside the particle, the volumetric component of the Minkowski force, represented by the second term of Equation (13), becomes negligible for a lossless dielectric medium. Consequently, the total electromagnetic force acting on such a material mainly arises from the first term of Equation (13), commonly referred to as the Helmholtz or surface Minkowski force [62]. There are several approaches to making this force negative for a lossless body [62]. It is always nonzero for a lossy body, such as plasmonic bodies, whereas the total force depends on both factors in Equation (13).

The bulk force in Equation (14) for lossy bodies leads to a positive bulk force because of its high dependency on the free current density. As discussed in Ref. [59], the repulsive force averaged overtime for plasmonic or lossy materials is primarily determined by the bulk contribution appearing in Eq. (13). Nevertheless, based on certain contemporary research studies [61–63], by optimizing both terms in Equation (13), it is feasible to obtain the total (overall) attractive force on a lossy body.

Equation (14) provides the mathematical interpretation of the individual terms appearing in Equation (13). In the present study, when the plasmonic nanorod undergoes an optical pulling effect, a force inversion occurs in the region where the induced current density of the plasmonic scatterer changes direction, as indicated by the bulk Minkowski force term in Equation (14). Figure 5(d) illustrates a distinct pulling region in which the plasmonic structure is subjected to an optical pulling force. This behavior is further confirmed in Fig. 5(b–f), where the arrow distributions of the induced current density clearly demonstrate the current reversal phenomenon. It is important to recognize that force distributions arising from bound charge and current interactions can take on both positive and negative magnitudes. Positive force densities generate repulsive or pushing effects, whereas negative force densities produce a backward-directed optical force responsible for the tractor-beam behavior along the (+z) axis [60]. Therefore, the occurrence of a pulling-type optical force in a plasmonic structure is primarily governed by a change in the direction of the induced current density. Another notable aspect of this phenomenon is the preserved alignment of the field distribution. Thus, the results suggest that the reversal of current density plays the key role in producing optical pulling forces for a plasmonic nanorod situated within the designed optical system. The reversal is also observed from Fig. 5(b-f) where the force acting on the plasmonic scatterer reverses its direction. We have discovered that this finding agrees with ours concerning the varying environmental scenarios.

### 3.3 Mechanism of optical pulling in a dielectric nanorod

To explain the origin of optical pulling force in a dielectric nanorod, one needs to present a theory based on multipole moments generated by an optical field. Eq. (15) may be employed for defining the time-average optical force arising in a dielectric particle due to the incidence of monochromatic optical radiation in the following way:

$$\langle \mathbf{F} \rangle = \mathbf{F}_{\mathrm{p}} + \mathbf{F}_{\mathrm{m}} + \mathbf{F}_{\mathrm{Qe}} + \mathbf{F}_{\mathrm{pm}} + \mathbf{F}_{\mathrm{Qep}} \tag{15}$$

Each term on the right-hand side of the equation corresponds, respectively, to the contributions arising from the electric dipole moment, magnetic dipole moment, and electric quadrupole moment, as well as the coupling between the electric and magnetic dipoles and the interaction between the electric dipole and electric quadrupole moments [64]. From their physical interpretation, these contributions may generally be separated into two categories: the first group originates directly from the incident electromagnetic beam, whereas the second group arises from interference-related interactions between different multipole moments [64–67]:

$$\mathbf{F} = \mathbf{F}_{\text{incident}} + \mathbf{F}_{\text{interference}} \tag{16}$$

Where,

$$\mathbf{F}_{\text{incident}} = \mathbf{F}_{\text{p}} + \mathbf{F}_{\text{m}} + \mathbf{F}_{\text{Qe}} \tag{17}$$

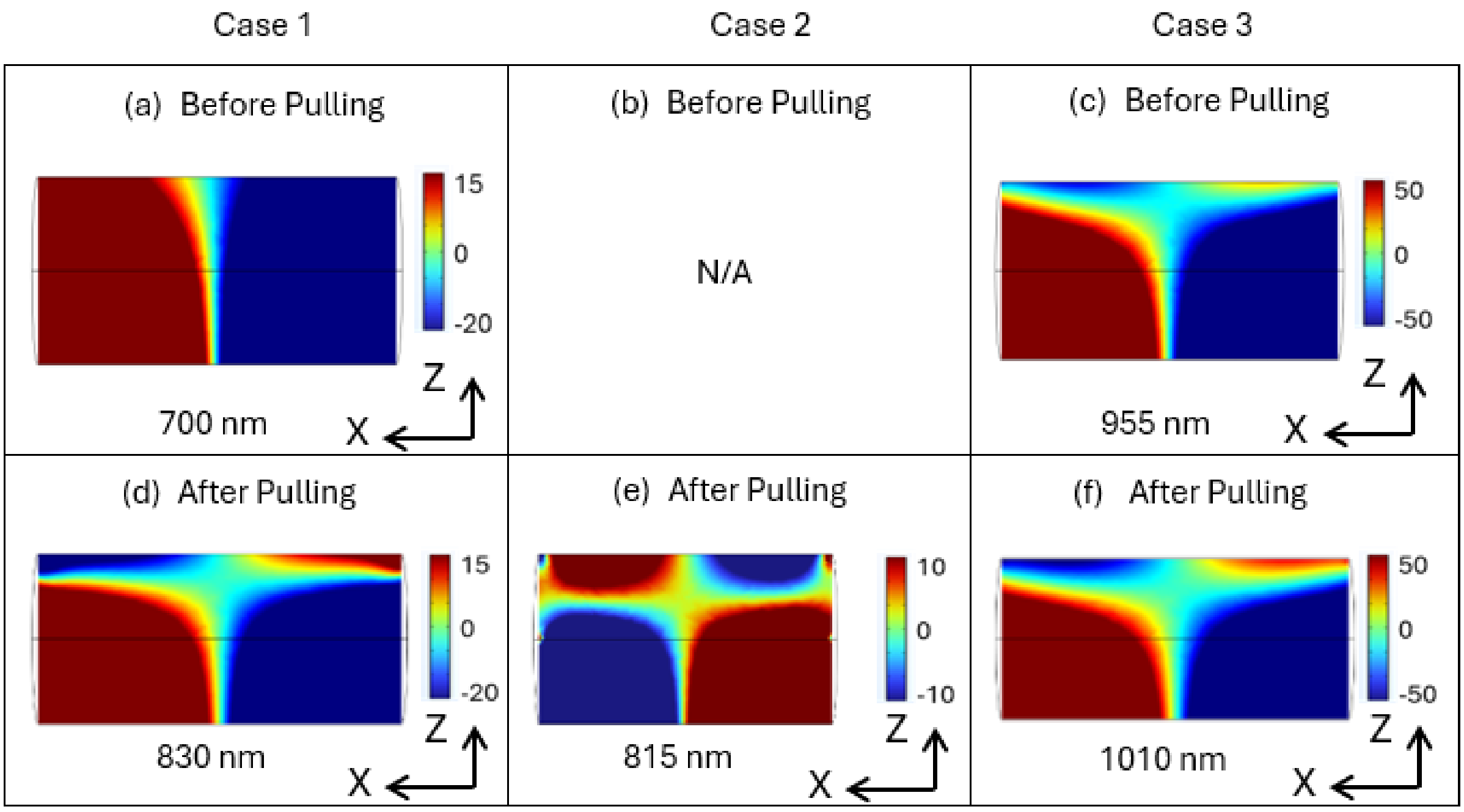

**Fig. 6.** illustrates the electric field (V/m) distributions around a dielectric nanorod for different incident wavelengths under three surrounding media. In case (1), the nanorod is placed in air and illuminated with incident beams of 700 nm and 830 nm, as shown in Figs. 6(a) and 6(d), respectively, using the proposed optical configuration. For case (2), the same nanorod is positioned in the identical setup but is partially immersed in water. The results for two different wavelengths are presented in Figs. 6(b) and 6(e). In Fig. 6(b), no significant field enhancement is observed, as the structure does not exhibit a distinct pulling region over the entire observation range. Finally, case (3) considers the nanorod fully immersed in water while being excited by incident wavelengths of 955 nm and 1010 nm, as shown in Figs. 6(c) and 6(f). These results highlight how the surrounding medium strongly influences the field distribution and optical response of the dielectric scatterer.

and,

$$\mathbf{F}_{\text{interference}} = \mathbf{F}_{\text{pm}} + \mathbf{F}_{\text{Qep}} \tag{18}$$

The previously introduced multipolar components may be expressed in the following form:

$$\mathbf{F}_{\text{p}} = \frac{1}{2}\,\text{Re}[(\nabla\mathbf{E}^{*})\bullet\,\mathbf{p}] \tag{19}$$

$$\mathbf{F}_{\text{m}} = \frac{1}{2}\,\text{Re}[(\nabla\mathbf{B}^{*})\bullet\,\mathbf{m}] \tag{20}$$

$$\mathbf{F}_{\mathrm{Qe}} = \frac{1}{4}\,\mathrm{Re}[(\nabla\nabla\mathbf{E}^{*})\bullet\,\mathbf{Q}_{e}] \tag{21}$$

$$\mathrm{F}_{\mathrm{pm}} = -\frac{k^{4}}{12\pi\epsilon_{0}c}\mathrm{Re}[\mathbf{p}\times\mathbf{m}^{*}] \tag{22}$$

$$\mathrm{F}_{\mathrm{Qep}} = -\frac{k^{4}}{40\pi\epsilon_{0}}\mathrm{Re}[\mathbf{Q}_{e}\bullet\,\mathbf{p}^{*}] \tag{23}$$

In the above expressions, $\mathbf{E}$ and $\mathbf{H}$ denote the applied electromagnetic field components corresponding to the electric and magnetic fields, respectively. The generated electric dipole, magnetic dipole, and electric quadrupole momentum are represented by $\mathbf{p} = \alpha_{\mathrm{e}}\mathbf{E}$, $\mathbf{m} = \alpha_{\mathrm{m}}\mathbf{H}$, and $\mathbf{Q}_{e} = \left(\frac{\gamma_{e}}{2}\right)(\nabla\mathbf{E} + \nabla\mathbf{E}^{\mathrm{T}})$. Here, $\alpha_{\mathrm{e}} = \frac{\mathrm{i}6\pi\epsilon_{0}\mathrm{a}_{1}}{\mathrm{k}^{3}}$, $\alpha_{\mathrm{m}} = \frac{\mathrm{i}6\pi\epsilon_{0}\mathrm{b}_{1}}{\mu_{0}\mathrm{k}^{3}}$, and $\gamma_{e} = \frac{\mathrm{i}40\pi\epsilon_{0}\mathrm{a}_{2}}{\mathrm{k}^{5}}$ indicate the complex-valued polarizability parameters, while $\mathrm{a}_{1}$, $\mathrm{a}_{2}$, $\mathrm{b}_{1}$ correspond to the scattering coefficients. The parameter "$\mathbf{K}$" represents the propagation wave vector. Using these definitions, Eq. (15) may alternatively be written in the following form:

$$\langle\mathbf{F}\rangle = \frac{1}{2}\,\mathrm{Re}[(\nabla\mathbf{E}^{*})\bullet\,\mathbf{p}] + \frac{1}{2}\,\mathrm{Re}[(\nabla\mathbf{B}^{*})\bullet\,\mathbf{m}] + \frac{1}{4}\,\mathrm{Re}[(\nabla\nabla\mathbf{E}^{*})\bullet\,\mathbf{Q}_{e}] - \frac{k^{4}}{12\pi\epsilon_{0}c}\mathrm{Re}[\mathbf{p}\times\mathbf{m}^{*}] - \frac{k^{4}}{40\pi\epsilon_{0}}\mathrm{Re}[\mathbf{Q}_{e}\bullet\,\mathbf{p}^{*}] \tag{24}$$

or,

$$\langle\mathbf{F}\rangle = \left(\frac{1}{2}\,\mathrm{Re}[(\nabla\mathbf{E}^{*})\bullet\,\mathbf{p}] + \frac{1}{2}\,\mathrm{Re}[(\nabla\mathbf{B}^{*})\bullet\,\mathbf{m}] + \frac{1}{4}\,\mathrm{Re}[(\nabla\nabla\mathbf{E}^{*})\bullet\,\mathbf{Q}_{e}]\right) - \left(\frac{k^{4}}{12\pi\epsilon_{0}c}\mathrm{Re}[\mathbf{p}\times\mathbf{m}^{*}] + \frac{k^{4}}{40\pi\epsilon_{0}}\mathrm{Re}[\mathbf{Q}_{e}\bullet\,\mathbf{p}^{*}]\right) \tag{25}$$

A single, unstructured, time-harmonic, plane-polarized laser beam, expressed in our case as $\mathrm{H} = \mathrm{H}_{0}\mathrm{e}^{\mathrm{j}kz}\hat{\mathrm{y}}$, can generate a spatial beating pattern along the propagation (z) direction. As a result, the optical intensity varies periodically along the z-axis. This spatial modulation gives rise to a net gradient force arising from the incident field contribution $\mathbf{F}_{\mathrm{incident}}$. This force may either be positive or negative. In this case, Eq. (25) can now be rewritten as:

$$\langle\mathbf{F}\rangle = \left(\frac{1}{2}\,\mathrm{Re}[(\nabla\mathbf{E}^{*})\bullet\,\mathbf{p}] + \frac{1}{2}\,\mathrm{Re}[(\nabla\mathbf{B}^{*})\bullet\,\mathbf{m}] + \frac{1}{4}\,\mathrm{Re}[(\nabla\nabla\mathbf{E}^{*})\bullet\,\mathbf{Q}_{e}] \pm |\mathrm{F}_{\mathrm{g}}|\right) - \left(\frac{k^{4}}{12\pi\epsilon_{0}c}\mathrm{Re}[\mathbf{p}\times\mathbf{m}^{*}] + \frac{k^{4}}{40\pi\epsilon_{0}}\mathrm{Re}[\mathbf{Q}_{e}\bullet\,\mathbf{p}^{*}]\right) \tag{26}$$

Here, $\mathrm{F}_{\mathrm{g}}$ denotes the gradient force defined in the equation above. As given in Eq. (26), the term $\varepsilon - |\,\mathrm{Fg}\,|$, together with the contribution from $\mathbf{F}_{\mathrm{interference}}$, leads to the emergence of an optical pulling force directed along the "+z" axis. In general, this pulling behavior is associated with ε, which corresponds to a negative gradient force component [68,69]. In contrast, the optical pushing force acting along the "-z" direction originates from the conventional contributions of $\mathbf{F}_{\text{incident}}$, as discussed in [70]. As a consequence, the periodic modulation of intensity along the z-direction can give rise to either attractive or repulsive optical forces acting on the nanorod. The development of a localized pole within the dielectric scatterer, observed in the proposed optical configuration, is shown in Fig. 6(d–f) for wavelengths of 830 nm, 815 nm, and 1010 nm corresponding to cases 1, 2, and 3, respectively. From the presented electric field distributions, it can be inferred that the optical pulling force experienced by the dielectric nanorod is mainly associated with interference between the electric dipole and quadrupole moments inside the structure. According to Eq. (26), the observed optical pulling phenomenon arises from the interaction between the electric dipole and electric quadrupole modes. In addition, the presence of the oppositely directed gradient-force component ($-|F_g|$) plays a significant role in this mechanism. The combined influence of these contributions ultimately leads to the appearance of an overall backward-directed optical force within the dielectric environment.

The electric-field profiles illustrated in Figs. 6(a) and 6(c) reveal a pronounced charge-separation response inside the dielectric nanorod positioned over the designed substrate. This behavior confirms the excitation of dipolar polarization modes within the structure. Such field characteristics are observed for the incident wavelength of 700 nm in Case 1, whereas a similar response appears at 955 nm for the Case 3 configuration. According to Eq. (26), the interplay between the electric dipole response and the positive gradient term (+|Fg|) generally results in an optical pushing effect. In contrast, the appearance of an optical pulling force is typically linked to the excitation of electric dipole–quadrupole interference together with the negative gradient component (−|Fg|). As a result, the interference-related force term $\mathbf{F}_{\text{interference}}$ can dominate over the conventional incident contribution $\mathbf{F}_{\text{incident}}$, leading to a net force directed opposite to the usual radiation pressure direction.

Ultimately, the optical tractor beam effect occurs due to the resulting dielectric object being pulled in the "+Z" direction. However, the optical pushing force is created when the electric dipole contribution and the positive gradient force (+ |Fg|) in the negative local direction dominates; hence a positive gradient force Fg is included in the net force given in Eq. (26). In this case, the optical pushing force arises when $\mathbf{F}_{\text{incident}}$ becomes greater than $\mathbf{F}_{\text{interference}}$.

## 4. Conclusion

In the present work, the concept of an optical system based on hyperbolic metamaterial (HMM) configurations has been experimentally validated as a feasible, straightforward, and versatile concept to induce optical pulling force on single and collective plasmonic, dielectric, and chiral nanorods. Through numerical simulations, it has been proven that the proposed design can induce optical pulling forces in three distinct media – complete air, half-water, and total water media. In the case of the collective nanorods, optical pulling forces have been successfully achieved within the wavelengths ranging between 800 and 1600 nm, depending on the medium. In the case of individual nanorods, optical pulling force can be induced for wavelengths of 700-1600 nm, depending on the nature of the nanorod. The underlying physics behind the optical pulling forces on each individual nanorod design, involving dipolar interference, current density switching, and multipole effects, has also been theoretically examined. To our knowledge, optical pulling force induction on all three categories of nanorods in all three media using a single hyperbolic framework has never been attempted before.

**Credential Authorship Contribution Statement**

**Md. Al Amin Chy** contributed to the work through conceptualization, data organization, formal analysis, investigation, methodological development, resource management, software implementation, validation, visualization, and preparation of the original manuscript draft, along with subsequent review and editing. **Md. Ajijul Haque Sakib Opu** participated in conceptualization, formal analysis, resource support, methodological planning, validation, visualization, and both drafting and revising the manuscript. **Md Fujael Ahmed** was involved in conceptualization, formal analysis, provision of resources, methodology design, validation, visualization, and preparation as well as revision of the manuscript. **Rashedul Albab** contributed through conceptualization, formal analysis, resource handling, methodological support, validation, visualization, and both writing and editing of the manuscript. **Md. Janibul Alam Soeb** provided supervision and guidance for the study, contributed to investigation and validation processes, assisted in visualization, managed project administration, and participated in reviewing and editing the manuscript.

**Competing Interests**

The authors state that they have no financial or personal conflicts of interest that could have affected or influenced the results presented in this study.